\documentclass[aip,rsi,reprint,floatfix,graphicx]{revtex4-1} 
\usepackage{graphicx}
\usepackage{dcolumn}
\usepackage{bm}
\usepackage{float}

\usepackage{amsmath}

\begin{document}


\title[]{Neural Network for 3D ICF Shell Reconstruction from Single Radiographs}

\thanks{Contributed paper to the Proceedings of the 23rd Topical Conference on High-Temperature Plasma Diagnostics, Santa Fe, NM, USA, May 31 - June 4, 2020. Rescheduled online, Dec. 14-17, 2020. Correspondence: (B.W.) bwolfe@lanl.gov, (Z.W.) zwang@lanl.gov.}

\author{Bradley T. Wolfe}
\affiliation{Los Alamos National Laboratory, Los Alamos, New Mexico 87545, USA}
\author{Zhizhong Han}
\affiliation{Department of Computer Science, University of Maryland, College Park, Maryland, 20742, USA}
\author{Jonathan S. Ben-Benjamin}
\affiliation{Los Alamos National Laboratory, Los Alamos, New Mexico 87545, USA}
\author{John L. Kline}
\affiliation{Los Alamos National Laboratory, Los Alamos, New Mexico 87545, USA}
\author{David S. Montgomery}
\affiliation{Los Alamos National Laboratory, Los Alamos, New Mexico 87545, USA}
\author{Elizabeth C. Merritt}
\affiliation{Los Alamos National Laboratory, Los Alamos, New Mexico 87545, USA}
\author{Paul A. Keiter}
\affiliation{Los Alamos National Laboratory, Los Alamos, New Mexico 87545, USA}
\author{Eric Loomis}
\affiliation{Los Alamos National Laboratory, Los Alamos, New Mexico 87545, USA}
\author{Brian M. Patterson}
\affiliation{Los Alamos National Laboratory, Los Alamos, New Mexico 87545, USA}
\author{Lindsey Kuettner}
\affiliation{Los Alamos National Laboratory, Los Alamos, New Mexico 87545, USA}
\author{Zhehui Wang}
\affiliation{Los Alamos National Laboratory, Los Alamos, New Mexico 87545, USA}

\date{\today}

\begin{abstract}
In inertial confinement fusion (ICF), X-ray radiography is a critical diagnostic for measuring implosion dynamics, which contains rich 3D information. Traditional methods for reconstructing 3D volumes from 2D radiographs, such as filtered backprojection, require radiographs from at least two different angles or lines of sight (LOS). In ICF experiments, space for diagnostics is limited and cameras that can operate on  the fast timescales are expensive to implement, limiting the number of projections that can be acquired. To improve the imaging quality as a result of this limitation, convolutional neural networks (CNN) have recently been shown to be capable of producing 3D models from visible light images or medical X-ray images rendered by volumetric computed tomography LOS (SLOS). We propose a CNN to reconstruct 3D ICF spherical shells from single radiographs. We also examine sensitivity of the 3D reconstruction to different illumination models using preprocessing techniques such as pseudo-flat fielding. To resolve the issue of the lack of 3D supervision, we show that training the CNN utilizing synthetic radiographs produced by known simulation methods allows for reconstruction of experimental data as long as the experimental data is similar to the synthetic data. We also show that the CNN allows for 3D reconstruction of shells that possess low mode asymmetries. Further comparisons of the 3D reconstructions with direct multiple LOS measurements are justified.
\end{abstract}

\maketitle

\section{Introduction}
In ICF, the drive asymmetry during implosion has been known to reduce neutron yield when compared to 1D simulations\cite{ LI201769, kyrala_delamater}. In double shells, low mode asymmetries from the ablator become exacerbated due to hydrodynamic instabilities such as Rayleigh–Taylor, Kelvin-Helmholtz, and Richtmyer–Meshkov. One technique to diagnose this asymmetry is by utilizing X-ray radiographic imaging during the experiment. This provides both challenges and opportunities in data interpretation which involve inverse problems. Asymmetry in the implosion can have three dimensional (3D) components while a radiograph is only a @D projection of this 3D information. 3D tomographic reconstruction is a well developed technique and is utilized in technologies such as X-ray computed tomography (CT) instruments. For this technique, a series of radiographs are collected from multiple angles which are then reconstructed using a variety of algorithms including filtered backprojection and iterative methods such as algebraic reconstruction techniques\cite{ COLSHER1977513, Willemink2019}. Furthermore, 3D reconstructions of time integrated implosions of deuterium filled capsules have been produced using a small number of views\cite{ doi:10.1063/1.4986652, doi:10.1063/1.4936319}. In ICF environments, room for diagnostics including X-ray imaging is scarce and the detectors that can operate on nanosecond to picosecond time scales are expensive to produce. Convolutional neural networks have been shown to effectively perform tasks such as segmentation, denoising, and feature recognition, whereas algorithms commonly used in libraries such as OpenCV\cite{ itseez2014theopencv} and scikit-image\cite{ van2014scikit}  rely on small parameter spaces that are manually selected. 
We utilize a convolutional neural network in order to reconstruct 3D spherical shell objects from X-ray radiographs generated from a single line of sight (SLOS). To resolve the issue of the lack of 3D supervision, synthetic data is generated to compensate for the very limited experimental data sets to successfully tune the network parameters. Data-driven 3D reconstruction from SLOS is a new concept for ICF applications. It will potentially allow us to reconstruct 3D objects using their 2D projections from any view angle, and therefore be able to compare with measurements directly, or to compare with simulations in unprecedented details.

This paper is divided into the following sections. We first describe the experimental setup used to produce the X-ray radiographs in Sec.\ \ref{Experimental Setup}. The architecture of the proposed convolutional neural network used is detailed in Sec.\ \ref{Neural Network Architecture}. In Sec.\ \ref{Synthetic Data Generation} the procedure for producing synthetic 2D-3D training samples is described. In Sec.\ \ref{Training} the neural network training process and associated hardware are described. Sec.\ \ref{results} shows results generate using the neural network. First, the 2D images rendered from both clean and noisy synthetic 3D shells are shown. Next, cross-sections of reconstructions from experimental data are shown. Then Fourier analysis is used to describe the effects of the neural network in frequency space. Finally, contours are extracted from cross-sections to show the utility of this network for Legendre mode analysis.

\section{Experimental Setup}
\label{Experimental Setup}
\begin{figure}
	\includegraphics[width=.43\textwidth]{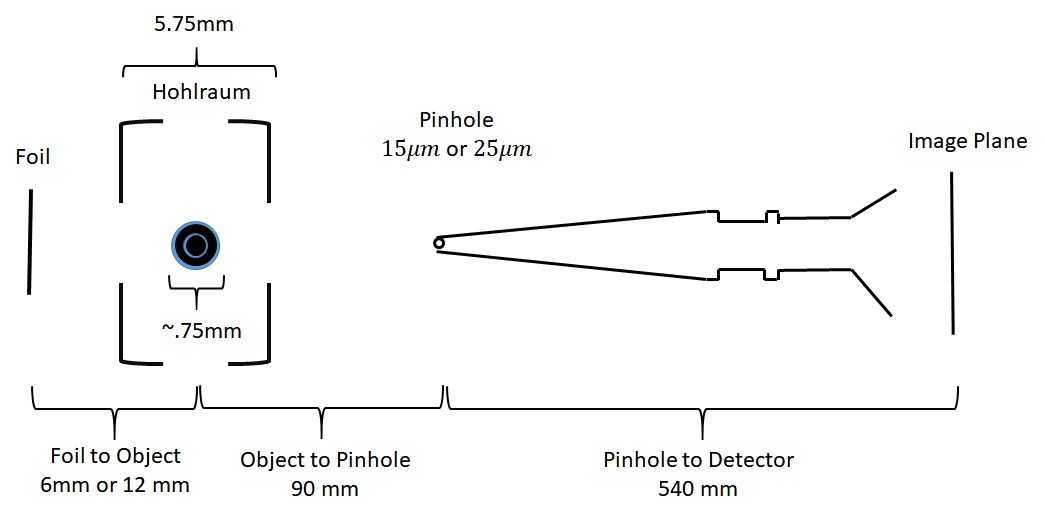}

	\caption{\label{experimentaldiagram} Schematic of the experimental setup. This includes the backlighter foil (far left), the hohlraum-capsule assembly (center left), and pinhole imager (right). The laser which impacts the foil to generate X-rays is not pictured and comes from the left.}%
\end{figure}

\begin{figure}
	\includegraphics[width=0.25\textwidth]{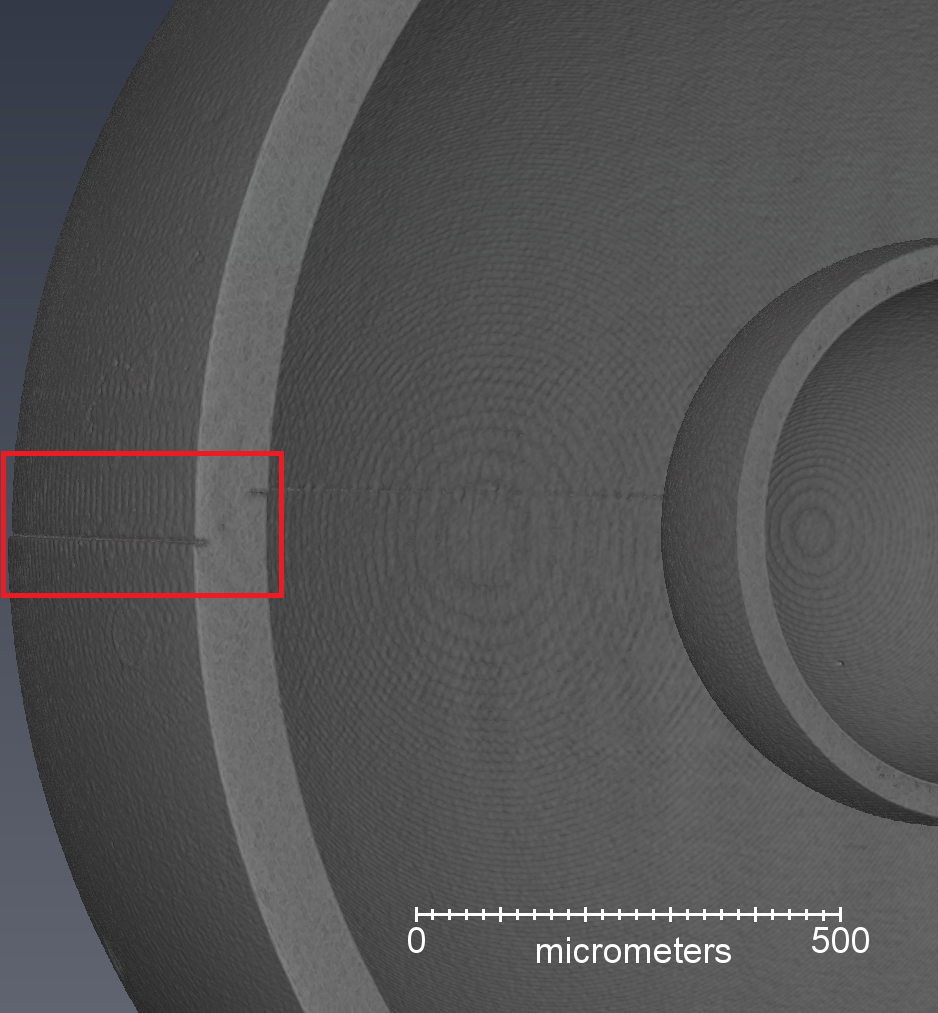}
	\caption{\label{3DRender} A high resolution 3D rendering from the reconstructed X-ray CT image of a double shell capsule. At this high resolution (1.5 $\mu$m voxel size) features such as the joint feature (shown in box) are visible.}
\end{figure}

Imaging data of double and single shell targets from NIF implosions are produced by using a Zr foil backlighter\cite{BARRIOS2013626, doi:10.1063/1.5086674} and a gated X-ray framing camera\cite{10.1117/12.513761}. The foil backlighter produces 16.3 keV X-rays which pass through a window in the hohlraum which eventually proceeds to the pinhole camera setup\cite{PhysRevLett.112.195001,doi:10.1063/1.5040995}. This pinhole camera utilizes microchannel plate strips, followed by a phosphor screen that is recorded by a CCD camera. The entire source-object-detector geometry of the experimental setup is shown in Fig.\ \ref{experimentaldiagram}. The spherical shell targets consist of an outer shell ablator and may contain an inner shell\cite{doi:10.1063/1.5086674, Cardenas}. Table \ref{table:Experimental Table} shows the materials contained in these targets. Fig.\ \ref{3DRender} shows a zoomed in rendering of the 3D reconstruction of the capsule used in shot N180918-001. The image was collected using a lab-based, micro-scale, X-ray CT instrument in which ~1261 radiographs were collected while the sample was rotated $360^{\circ}$ which were reconstructed and rendered.

\begin{table}[]
\scalebox{0.9}{
	\begin{tabular}{|l|l|l|l|}
		Shot Number & Inner Shell                 & Outer Shell   & Capsule Fill  \\ \hline
		N170222-003 & 40$\mu$m Si$O_2$           & 106$\mu$m Al & N/A           \\
		N170322-001 & N/A                         & 106$\mu$m Al & 35 mg/${cm}^3$ foam \\
		N171016-001 & 40$\mu$m Si$O_2$                     & 106$\mu$m Al & deuterium     \\
		N180321-003 & N/A                         & 106$\mu$m Al & N/A           \\
		N180522-002 & N/A                         & 120$\mu$m Al & N/A           \\
		N180731-002 & 40$\mu$m Si$O_2$(Ge doped) & 106$\mu$m Al & deuterium     \\
		N180918-001 & 40$\mu$m Si$O_2$(Ge doped) & 106$\mu$m Al & deuterium    
	\end{tabular}
}
\caption{Configurations of materials in targets. The shells that contain a doping are 1.5\% germanium.}
\label{table:Experimental Table}
\end{table}

\section{Neural Network Architecture} 
\label{Neural Network Architecture}
\begin{figure}[H]
	\includegraphics[width=.45\textwidth]{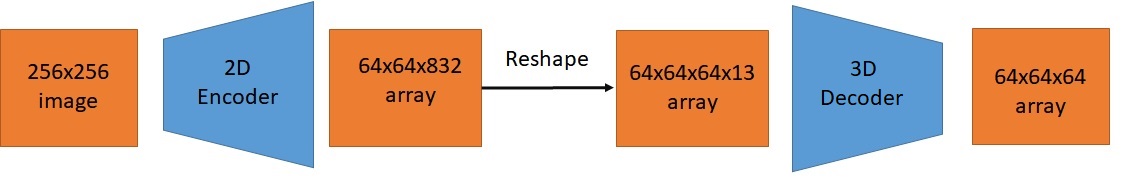}
	
	\caption{\label{arch} The architecture of the network given a 256-by-256 input and a 64-by-64-by-64 output. This can be amended to higher input and output sizes by changing the number of convolutional filters used in the final layer of the 2D encoder.}%
\end{figure}

The convolutional neural network proposed for this task is based upon the encoder structure from transformable bottleneck networks\cite{olszewski2019transformable, TBN} (TBN). TBN's are used in the task of novel view synthesis for objects given multiple input views, while generating an internal volumetric render. In the case of X-ray radiography, the volumetric reconstruction is the more important piece of information, since producing a different projection of the object is rather simple. This means that the decoder structure of the TBN can be neglected. Furthermore, the use of the resampling layer is unnecessary since only a SLOS is being used. This leaves a 2D encoder, a reshaping layer, and a 3D decoder. In total these modules contain 11,962,313 trainable parameters. Fig \ref{arch} shows the shape of an image as it is passed through the convolutional neural network. The 832 dimension is due to the fact that the 2D encoder's final convolutional layer contains 832 filters. For different input image sizes the number of features can be changed so that the number is divisible by the 3D array dimensions.

\section{Synthetic Data Generation}
\label{Synthetic Data Generation}
To train the model, ground truth 3D volumetric objects are required along with the corresponding 2D projections. However, producing large sets of images in an experimental setting is prohibitively expensive and an accurate 3D ground truth is unknown. The inverse of this problem however is much better understood. Given a 3D object with known material properties, a well defined source-detector geometry, and well known source properties, projections can be easily produced. This is due to the Beer-Lambert law which calculates the transmittance of light through an object as:
\begin{equation}
T := \frac{I}{I_0}= e^{-\int\mu ds} 
\end{equation}

Where $\mu$ is the linear attenuation of the material which can vary in different regions of the object and the integral is a line integral from the source to a position on the detector. The linear attenuation is calculated by using the mass attenuation coefficients of the materials and the material density of the object at room temperature. X-ray mass attenuation coefficients for different materials are tabulated on the NIST Standard reference Database 8\cite{NistAttenuation}. 

While the value of $\mu$ is related to material properties and the energy of the incident photons, the line integral is determined by the geometry of the object and the source-detector geometry. Since this operation must be performed for each pixel of the detector, a GPU based tomography library is used to speed up these calculations. The Tomographic Iterative GPU-based Reconstruction Toolbox (TIGRE)\cite{Biguri_2016, BIGURI202052} provides this functionality and allows for a variety of source-detector-object geometries.

\begin{figure}
	\includegraphics[width=0.45\textwidth]{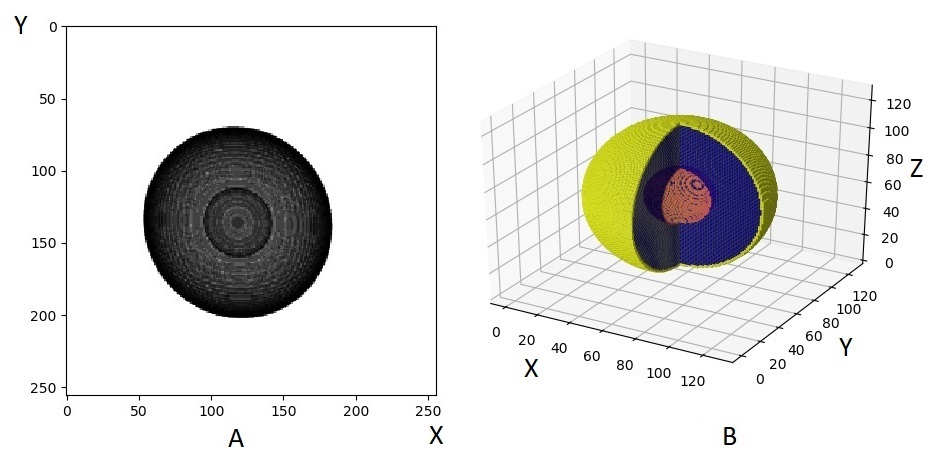}
	
	\caption{\label{syntheticdata} A) A projection of an automatically generated object using TIGRE. B) The automatically generated object in 3D. A quarter of the object has been removed to aid the reader in visualizing the interior structure. Colors indicate various materials.}
\end{figure}

Given a model for attenuation and the ability to calculate the line integral through the object, synthetic SLOS radiographic images are automatically generated by randomly sampling source parameters and 3D object properties. Furthermore, the outer radius of the object is selected from a range of 0.15 mm and 1.5 mm, the outer shell thickness is drawn from a range of 1\%-30\% of the outer shell radius, and a skew transform is applied to add a variable amount of asymmetry to the targets. Using TIGRE, synthetic SLOS radiographic images can be generated as the projections of these objects, which are regarded as the 3D ground truth, to train the CNN. By multiplying the calculated transmission array by a constant intensity profile radiographs can be produced under different illumination conditions. Fig.\ \ref{syntheticdata} shows an example of the generated data from the synthetic model.

We use an Additive Gaussian Noise Model where the spread of the distribution varies and the Gaussian is sampled over integers. This is done since the image is treated as counts detected with some variation due to noise. First the standard deviation $\sigma$ was selected using a uniformly distributed distribution on integers 10 through 30. Then a distribution is defined between the values -100 and 100 where the probability of selecting any given integer is given by a Gaussian distribution with standard deviation $\sigma$. Giving a maximum and minimum value for the noise offset restricts the number of counts which makes the noisy generated data well behaved.

\begin{figure}
	\includegraphics[width=0.35\textwidth]{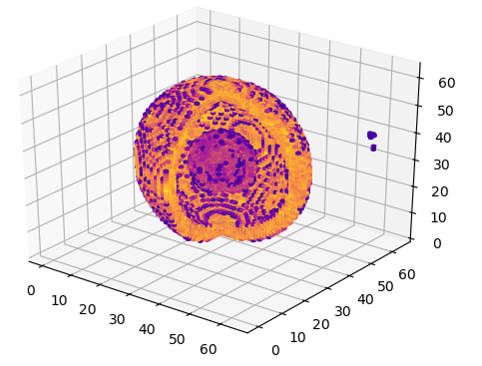}
	
	\caption{\label{clean} A 3D reconstruction from synthetic data using a network trained on data without noise. The reconstruction shows the presence of an inner shell and outer shell. In this plot background voxel values were set to zero by using a simple threshold. Colors are used to indicate various materials. The purple voxels found on the shells and off to the right have values that are above the used threshold.}%
\end{figure}

\section{Training}
\label{Training}
The CNN is trained using stochastic gradient descent on the mean squared error (MSE) between the output of the network ($f(X)$) and the 3D ground truth ($Y$), where the average is taken over the number ($N$) of voxels (3D pixels). 
\begin{equation}
	MSE = \frac{1}{N}||f(X) - Y||_2^2
\end{equation} 

Generated datasets with two thousand pairs of projections and 3D objects are used as a training set to optimize the parameters of the network. Two separate datasets, one with a noise background and one without. These models are trained for a large number of epochs ($~$300 epochs). The optimization was done using a batch size of 10 and a learning rate of 0.01. The model efficacy during training is measured using an independently generated set of 300 pairs. 

\section{Results}
\label{results}
\begin{figure}
	\includegraphics[width=0.35\textwidth]{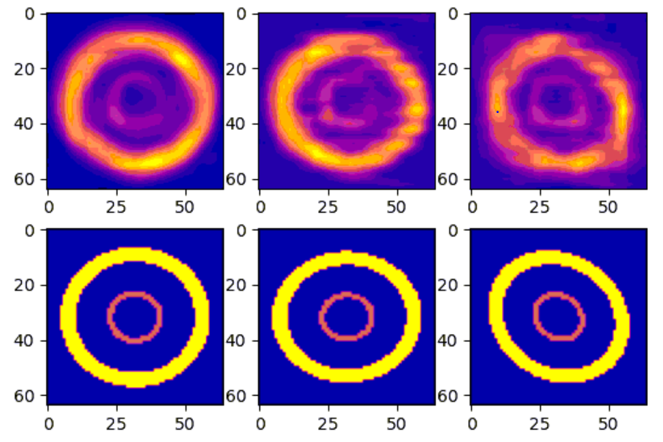}
	
	\caption{\label{noisy} Cross-sections of a reconstructed 3D spherical shell from synthetic data using a network trained on images with noise. Images from left to right are cross-sections of the reconstruction taken from the center on orthogonal planes. The top row shows cross-sections of the reconstruction produced by the network. The bottom row shows the ground truth from the synthetic X-ray CT images.}%
\end{figure}

Figs. \ref{clean} and \ref{noisy} visualize the reconstruction produced from the two trained models described in the prior section. In both cases the reconstruction contains both an inner and outer shell. These reconstructions are both reasonable since they share a common geometry with the original objects and are accurate when compared numerically. The noisy data is shown as cross-sections since the background voxels are more difficult to extract from the object. The network is able to reconstruct 3D shells regardless of the size of the object and on varied source energies and intensities. 

While the synthetic data is similar to the experimental data in the fact that the experimental data possesses an inner shell and an outer shell, there are characteristics that the experimental images possess that the synthetic images do not. One example of this is the illumination of the object. All of the synthetic data used to train the networks have a constant illumination across the image. In the experimental settings the backlighter provides a non-uniform illumination which causes regions of images to be brighter or dimmer without the density of the object materials contributing to this effect. This causes an issue since convolutional neural networks perform better on inputs that are more similar to the training data. Either the synthetic data needs to account for this or the experimental data can be preprocessed. In image processing this illumination problem is typically solved by using a flat field correction. This is done typically by dividing the image by a flat field image. In the case of ICF imaging a true flat field cannot be obtained since this would require taking an image of the backlighter under the same conditions without the object and these conditions would be difficult to reproduce. We instead use a pseudo-flat field correction\cite{flatfield} to preprocess the experimental images. In pseudo-flat fielding, the flat field is produced by a median filter of the image followed by a Gaussian filter. For the median filter a large disk filter is used and for the Gaussian filter a large sigma value is used. This captures larger trends in the illumination of the image. 

\begin{figure}
	\includegraphics[width=0.35\textwidth]{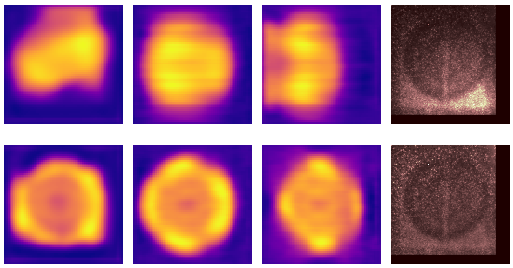}
	
	\caption{\label{ex_res} (Top) Cross-sections of the reconstruction from the experimental image (right). (Bottom) Cross-sections of the reconstruction from the pseudo-flatfield corrected image (right).}%
\end{figure}

Since the experimentally produced images contain noise, the neural network trained on noisy synthetic data is used for reconstruction. Fig.\ \ref{ex_res} shows the results of application of the network on the preprocessed experimental image and raw experimental image. In the case of the raw image the non-uniform illumination in the images causes the reconstruction to miss the key features in the experimental image such as the shells. By flattening the image using pseudo-flatfielding the network then picks up the key details in the image. The reconstruction even contains an inner shell which is typically difficult to detect in these images. Similarly to the noisy synthetic data, the background voxels are difficult to subtract since the resulting array is less sharp when compared to the noiseless synthetic data.   

\begin{figure}
	\includegraphics[width=0.30\textwidth]{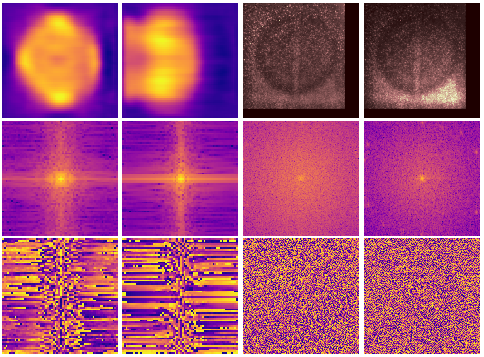}
	\caption{\label{fourier}  (Top) Image analyzed using Fourier Transform. (Middle) Magnitude spectrums from cross-sections of reconstructions have more less high frequency contributions when compared to the experimental images. (Bottom)The phase spectra from cross-sectional reconstructions are more structured than the experimental counterparts.}%
\end{figure}

While convolutional neural networks are able to produce impressive results, understanding these models can prove to be quite difficult. Since convolutional neural networks are known to identify structures in images through spatial filters, Fourier analysis provides a manner to analyze these results. The two dimensional Fourier transform produces frequencies that are present in images. The transformed image is typically displayed in the form
\begin{equation}
\hat{I}[i,j] = A[i,j]e^{i\phi[i,j]}
\end{equation} 

where $A$ is the magnitude spectrum and $\phi$ is the phase spectrum which are both real valued and can thus be represented as images themselves. Values closer to the center of these images correspond to smaller frequency values, while values further away correspond to higher frequency values. In Fig.\ \ref{fourier} the Fourier transforms of different types of images used in this work are shown. The phase specta found in the cross-section of the reconstruction possess more structure than the phase spectra that originate from the experimental images. In images the phase provides information of features such as edges. The improved signal-to-noise of the cross-sectional image allows for visualization of this structured phase. In the magnitude spectra of cross-section, pixels closer the center have higher average values relative to those away from the center. This is due to the fact that the images are low noise and smooth which correspond to low frequency components. Therefore, one benefit of the network, which is that it produces low noise reconstruction from noisy inputs, are now easier to analyze.

\begin{figure}
	\includegraphics[width=0.38\textwidth]{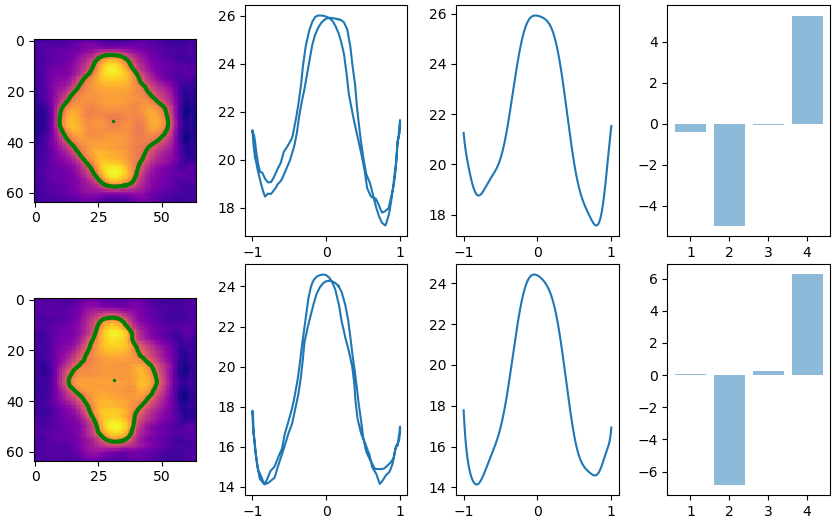}
	\caption{\label{modes} Legendre mode fitting on cross-sections. The cross-section in the bottom row is of the same object as the top row but at a later time. (Left) Contour detection of cross-section, (Middle Left) Contour radius w.r.t $cos\theta$, (Middle Right) Legendre Curve fit using N=15. (Right) coefficients of low mode coeffients, Even modes are prominent while odd modes contribute less. The radius $r(\theta)$ is also lower in the later cross-section.}%
\end{figure}

Since the cross-sectional images are low in noise, the shells are easier to extract using contour based algorithms. From thes images, Legendre mode analysis can be completed. A marching squares algorithm found in scikit-image is used to extract the image contours. By locating the center of the object, this contour was parametrized as a function of angle $r(\theta)$. A fit is applied to the data using Legendre polynomials,
\begin{equation}
	r(\theta) = \sum_{n=0}^{n=N}c_nP_n(cos\theta)
\end{equation}   
where $P_n$ is the $n^{th}$ degree Legendre polynomial. Fig.\ \ref{modes} shows the process of contour detection being applied on a cross-section of the reconstruction in order to measure Legendre coefficients.

\section{Conclusion}
We demonstrated the use of a convolutional neural network for 3D reconstruction of ICF capsules from single line of sight radiographs. This technique shows the benefit of leveraging synthetic radiographs in order account for the unknown shape of the capsule during implosion and the small amount of data that can be produced in experimental conditions. Furthermore this technique is able to be used on asymmetric objects which commonly appear in ICF environments. This neural network model is able to capture internal structure of the capsules such as the inner shell which is normally difficult to detect using conventional means. Further expansion on the method is warranted to generate higher resolution objects, synthetic models which better represent experimental conditions, and support multiple line-of-sight measurements. 

\section{Acknowledgment}
The authors would like to thank Pawel Kozlowski (Los Alamos National Laboratory) for providing code for the pseudo-flatfielding preprocessing.


\section*{References}
\nocite{*}

\bibliography{reference}

\begin{thebibliography}{22}%
\makeatletter
\providecommand \@ifxundefined [1]{%
 \@ifx{#1\undefined}
}%
\providecommand \@ifnum [1]{%
 \ifnum #1\expandafter \@firstoftwo
 \else \expandafter \@secondoftwo
 \fi
}%
\providecommand \@ifx [1]{%
 \ifx #1\expandafter \@firstoftwo
 \else \expandafter \@secondoftwo
 \fi
}%
\providecommand \natexlab [1]{#1}%
\providecommand \enquote  [1]{``#1''}%
\providecommand \bibnamefont  [1]{#1}%
\providecommand \bibfnamefont [1]{#1}%
\providecommand \citenamefont [1]{#1}%
\providecommand \href@noop [0]{\@secondoftwo}%
\providecommand \href [0]{\begingroup \@sanitize@url \@href}%
\providecommand \@href[1]{\@@startlink{#1}\@@href}%
\providecommand \@@href[1]{\endgroup#1\@@endlink}%
\providecommand \@sanitize@url [0]{\catcode `\\12\catcode `\$12\catcode
  `\&12\catcode `\#12\catcode `\^12\catcode `\_12\catcode `\%12\relax}%
\providecommand \@@startlink[1]{}%
\providecommand \@@endlink[0]{}%
\providecommand \url  [0]{\begingroup\@sanitize@url \@url }%
\providecommand \@url [1]{\endgroup\@href {#1}{\urlprefix }}%
\providecommand \urlprefix  [0]{URL }%
\providecommand \Eprint [0]{\href }%
\providecommand \doibase [0]{http://dx.doi.org/}%
\providecommand \selectlanguage [0]{\@gobble}%
\providecommand \bibinfo  [0]{\@secondoftwo}%
\providecommand \bibfield  [0]{\@secondoftwo}%
\providecommand \translation [1]{[#1]}%
\providecommand \BibitemOpen [0]{}%
\providecommand \bibitemStop [0]{}%
\providecommand \bibitemNoStop [0]{.\EOS\space}%
\providecommand \EOS [0]{\spacefactor3000\relax}%
\providecommand \BibitemShut  [1]{\csname bibitem#1\endcsname}%
\let\auto@bib@innerbib\@empty
\bibitem [{\citenamefont {Li}(2017)}]{LI201769}%
  \BibitemOpen
  \bibfield  {author} {\bibinfo {author} {\bibfnamefont {Y.}~\bibnamefont
  {Li}},\ }\href {\doibase https://doi.org/10.1016/j.mre.2016.12.001}
  {\bibfield  {journal} {\bibinfo  {journal} {Matter and Radiation at
  Extremes}\ }\textbf {\bibinfo {volume} {2}},\ \bibinfo {pages} {69 }
  (\bibinfo {year} {2017})},\ \bibinfo {note} {special Issue on Laser Fusion
  (II)}\BibitemShut {NoStop}%
\bibitem [{\citenamefont {Kyrala}(2005)}]{kyrala_delamater}%
  \BibitemOpen
  \bibfield  {author} {\bibinfo {author} {\bibfnamefont {G.~A.}\ \bibnamefont
  {Kyrala}},\ }\href {\doibase 10.1017/S0263034605050330} {\bibfield  {journal}
  {\bibinfo  {journal} {Laser and Particle Beams}\ }\textbf {\bibinfo {volume}
  {23}},\ \bibinfo {pages} {187–192} (\bibinfo {year} {2005})}\BibitemShut
  {NoStop}%
\bibitem [{\citenamefont {Colsher}(1977)}]{COLSHER1977513}%
  \BibitemOpen
  \bibfield  {author} {\bibinfo {author} {\bibfnamefont {J.~G.}\ \bibnamefont
  {Colsher}},\ }\href {\doibase https://doi.org/10.1016/S0146-664X(77)80014-2}
  {\bibfield  {journal} {\bibinfo  {journal} {Computer Graphics and Image
  Processing}\ }\textbf {\bibinfo {volume} {6}},\ \bibinfo {pages} {513 }
  (\bibinfo {year} {1977})}\BibitemShut {NoStop}%
\bibitem [{\citenamefont {Willemink}\ and\ \citenamefont
  {B.No{\"e}l.}(2019)}]{Willemink2019}%
  \BibitemOpen
  \bibfield  {author} {\bibinfo {author} {\bibfnamefont {M.~J.}\ \bibnamefont
  {Willemink}}\ and\ \bibinfo {author} {\bibfnamefont {P.}~\bibnamefont
  {B.No{\"e}l.}},\ }\href {\doibase 10.1007/s00330-018-5810-7} {\bibfield
  {journal} {\bibinfo  {journal} {European Radiology}\ }\textbf {\bibinfo
  {volume} {29}},\ \bibinfo {pages} {2185} (\bibinfo {year}
  {2019})}\BibitemShut {NoStop}%
\bibitem [{\citenamefont {Volegov}(2017)}]{doi:10.1063/1.4986652}%
  \BibitemOpen
  \bibfield  {author} {\bibinfo {author} {\bibfnamefont {P.~L.}\ \bibnamefont
  {Volegov}},\ }\href {\doibase 10.1063/1.4986652} {\bibfield  {journal}
  {\bibinfo  {journal} {Journal of Applied Physics}\ }\textbf {\bibinfo
  {volume} {122}},\ \bibinfo {pages} {175901} (\bibinfo {year} {2017})},\
  \Eprint {http://arxiv.org/abs/https://doi.org/10.1063/1.4986652}
  {https://doi.org/10.1063/1.4986652} \BibitemShut {NoStop}%
\bibitem [{\citenamefont {Volegov}(2015)}]{doi:10.1063/1.4936319}%
  \BibitemOpen
  \bibfield  {author} {\bibinfo {author} {\bibfnamefont {P.~L.}\ \bibnamefont
  {Volegov}},\ }\href {\doibase 10.1063/1.4936319} {\bibfield  {journal}
  {\bibinfo  {journal} {Journal of Applied Physics}\ }\textbf {\bibinfo
  {volume} {118}},\ \bibinfo {pages} {205903} (\bibinfo {year} {2015})},\
  \Eprint {http://arxiv.org/abs/https://doi.org/10.1063/1.4936319}
  {https://doi.org/10.1063/1.4936319} \BibitemShut {NoStop}%
\bibitem [{its(2014)}]{itseez2014theopencv}%
  \BibitemOpen
  \href@noop {} {\emph {\bibinfo {title} {The OpenCV Reference Manual}}},\
  \bibinfo {organization} {Itseez},\ \bibinfo {edition} {2nd}\ ed. (\bibinfo
  {year} {2014})\BibitemShut {NoStop}%
\bibitem [{\citenamefont {der Walt}(2014)}]{van2014scikit}%
  \BibitemOpen
  \bibfield  {author} {\bibinfo {author} {\bibfnamefont {V.}~\bibnamefont {der
  Walt}},\ }\href@noop {} {\bibfield  {journal} {\bibinfo  {journal} {PeerJ}\
  }\textbf {\bibinfo {volume} {2}},\ \bibinfo {pages} {e453} (\bibinfo {year}
  {2014})}\BibitemShut {NoStop}%
\bibitem [{\citenamefont {Barrios}(2013)}]{BARRIOS2013626}%
  \BibitemOpen
  \bibfield  {author} {\bibinfo {author} {\bibfnamefont {M.}~\bibnamefont
  {Barrios}},\ }\href {\doibase https://doi.org/10.1016/j.hedp.2013.05.018}
  {\bibfield  {journal} {\bibinfo  {journal} {High Energy Density Physics}\
  }\textbf {\bibinfo {volume} {9}},\ \bibinfo {pages} {626 } (\bibinfo {year}
  {2013})}\BibitemShut {NoStop}%
\bibitem [{\citenamefont {Merritt}(2019)}]{doi:10.1063/1.5086674}%
  \BibitemOpen
  \bibfield  {author} {\bibinfo {author} {\bibfnamefont {E.~C.}\ \bibnamefont
  {Merritt}},\ }\href {\doibase 10.1063/1.5086674} {\bibfield  {journal}
  {\bibinfo  {journal} {Physics of Plasmas}\ }\textbf {\bibinfo {volume}
  {26}},\ \bibinfo {pages} {052702} (\bibinfo {year} {2019})}\BibitemShut
  {NoStop}%
\bibitem [{\citenamefont {Oertel}(2004)}]{10.1117/12.513761}%
  \BibitemOpen
  \bibfield  {author} {\bibinfo {author} {\bibfnamefont {J.~A.}\ \bibnamefont
  {Oertel}},\ }\bibinfo {organization} {International Society for Optics and
  Photonics}\ (\bibinfo  {publisher} {SPIE},\ \bibinfo {year} {2004})\ pp.\
  \bibinfo {pages} {214 -- 222}\BibitemShut {NoStop}%
\bibitem [{\citenamefont {Rygg}(2014)}]{PhysRevLett.112.195001}%
  \BibitemOpen
  \bibfield  {author} {\bibinfo {author} {\bibfnamefont {J.~R.}\ \bibnamefont
  {Rygg}},\ }\href {\doibase 10.1103/PhysRevLett.112.195001} {\bibfield
  {journal} {\bibinfo  {journal} {Phys. Rev. Lett.}\ }\textbf {\bibinfo
  {volume} {112}},\ \bibinfo {pages} {195001} (\bibinfo {year}
  {2014})}\BibitemShut {NoStop}%
\bibitem [{\citenamefont {Loomis}(2018)}]{doi:10.1063/1.5040995}%
  \BibitemOpen
  \bibfield  {author} {\bibinfo {author} {\bibfnamefont {E.~N.}\ \bibnamefont
  {Loomis}},\ }\href {\doibase 10.1063/1.5040995} {\bibfield  {journal}
  {\bibinfo  {journal} {Physics of Plasmas}\ }\textbf {\bibinfo {volume}
  {25}},\ \bibinfo {pages} {072708} (\bibinfo {year} {2018})},\ \Eprint
  {http://arxiv.org/abs/https://doi.org/10.1063/1.5040995}
  {https://doi.org/10.1063/1.5040995} \BibitemShut {NoStop}%
\bibitem [{\citenamefont {Cardenas}()}]{Cardenas}%
  \BibitemOpen
  \bibfield  {author} {\bibinfo {author} {\bibfnamefont {T.}~\bibnamefont
  {Cardenas}},\ }\href@noop {} {\bibfield  {journal} {\bibinfo  {journal}
  {Fusion Science and Technology}\ }\textbf {\bibinfo {volume} {73}},\ \bibinfo
  {pages} {344}}\BibitemShut {NoStop}%
\bibitem [{\citenamefont {Olszewski}(2019)}]{olszewski2019transformable}%
  \BibitemOpen
  \bibfield  {author} {\bibinfo {author} {\bibfnamefont {K.}~\bibnamefont
  {Olszewski}},\ }\href@noop {} {\enquote {\bibinfo {title} {Transformable
  bottleneck networks},}\ } (\bibinfo {year} {2019}),\ \Eprint
  {http://arxiv.org/abs/1904.06458} {arXiv:1904.06458 [cs.CV]} \BibitemShut
  {NoStop}%
\bibitem [{\citenamefont {Olszewski}()}]{TBN}%
  \BibitemOpen
  \bibfield  {author} {\bibinfo {author} {\bibfnamefont {K.}~\bibnamefont
  {Olszewski}},\ }\href@noop {} {\enquote {\bibinfo {title} {Transformable
  bottleneck networks},}\ }\Eprint
  {http://arxiv.org/abs/https://github.com/kyleolsz/TB-Networks}
  {https://github.com/kyleolsz/TB-Networks} \BibitemShut {NoStop}%
\bibitem [{\citenamefont {Berger}(2010)}]{NistAttenuation}%
  \BibitemOpen
  \bibfield  {author} {\bibinfo {author} {\bibfnamefont {M.~J.}\ \bibnamefont
  {Berger}},\ }\href {\doibase 10.18434/T48G6X} {\enquote {\bibinfo {title}
  {Xcom: Photon cross sections database},}\ } (\bibinfo {year}
  {2010})\BibitemShut {NoStop}%
\bibitem [{\citenamefont {Biguri}(2016)}]{Biguri_2016}%
  \BibitemOpen
  \bibfield  {author} {\bibinfo {author} {\bibfnamefont {A.}~\bibnamefont
  {Biguri}},\ }\href {\doibase 10.1088/2057-1976/2/5/055010} {\bibfield
  {journal} {\bibinfo  {journal} {Biomedical Physics {\&} Engineering Express}\
  }\textbf {\bibinfo {volume} {2}},\ \bibinfo {pages} {055010} (\bibinfo {year}
  {2016})}\BibitemShut {NoStop}%
\bibitem [{\citenamefont {Biguri}(2020)}]{BIGURI202052}%
  \BibitemOpen
  \bibfield  {author} {\bibinfo {author} {\bibfnamefont {A.}~\bibnamefont
  {Biguri}},\ }\href {\doibase https://doi.org/10.1016/j.jpdc.2020.07.004}
  {\bibfield  {journal} {\bibinfo  {journal} {Journal of Parallel and
  Distributed Computing}\ }\textbf {\bibinfo {volume} {146}},\ \bibinfo {pages}
  {52 } (\bibinfo {year} {2020})}\BibitemShut {NoStop}%
\bibitem [{\citenamefont {{Pawel Kozlowski }}()}]{flatfield}%
  \BibitemOpen
  \bibfield  {author} {\bibinfo {author} {\bibnamefont {{Pawel Kozlowski }}},\
  }\href@noop {} {\enquote {\bibinfo {title} {Pseudo-flatfielding python
  code},}\ }\bibinfo {howpublished} {(private communication)}\BibitemShut
  {NoStop}%
\bibitem [{\citenamefont {Field}(2014)}]{doi:10.1063/1.4890395}%
  \BibitemOpen
  \bibfield  {author} {\bibinfo {author} {\bibfnamefont {J.}~\bibnamefont
  {Field}},\ }\href {\doibase 10.1063/1.4890395} {\bibfield  {journal}
  {\bibinfo  {journal} {Review of Scientific Instruments}\ }\textbf {\bibinfo
  {volume} {85}},\ \bibinfo {pages} {11E503} (\bibinfo {year} {2014})},\
  \Eprint {http://arxiv.org/abs/https://doi.org/10.1063/1.4890395}
  {https://doi.org/10.1063/1.4890395} \BibitemShut {NoStop}%
\bibitem [{\citenamefont {Varnum}(2000)}]{PhysRevLett.84.5153}%
  \BibitemOpen
  \bibfield  {author} {\bibinfo {author} {\bibfnamefont {W.~S.}\ \bibnamefont
  {Varnum}},\ }\href {\doibase 10.1103/PhysRevLett.84.5153} {\bibfield
  {journal} {\bibinfo  {journal} {Phys. Rev. Lett.}\ }\textbf {\bibinfo
  {volume} {84}},\ \bibinfo {pages} {5153} (\bibinfo {year}
  {2000})}\BibitemShut {NoStop}%
\end{thebibliography}%
\end{document}